# Safety Challenges and Solutions in Mobile Social Networks

Yashar Najaflou, Behrouz Jedari, Feng Xia, *Senior Member, IEEE*,
Laurence T. Yang, and Mohammad S. Obaidat, *Fellow, IEEE*

*Abstract*—Mobile social networks (MSNs) are specific types of social media which consolidate the ability of omnipresent connection for mobile users/devices to share user-centric data objects among interested users. Taking advantage of the characteristics of both social networks and opportunistic networks (OppNets), MSNs are capable of providing an efficient and effective mobile environment for users to access, share, and distribute data. However, the lack of a protective infrastructure in these networks has turned them into convenient targets for various perils. This is the main impulse why MSNs carry disparate and intricate safety concerns and embrace divergent safety challenging problems. In this paper, we aim to provide a clear categorization on safety challenges and a deep exploration over some recent solutions in MSNs. This work narrows the safety challenges and solution techniques down from OppNets and delay-tolerant networks to MSNs with the hope of covering all the work proposed around security, privacy, and trust in MSNs. To conclude, several major open research issues are discussed, and future research directions are outlined.

## I. Introduction

**T**HE INCREASE in the number of mobile devices has enabled users to be ubiquitously connected through wireless and mobile communications technologies. However, unlike conventional mobile ad hoc networks, persistent connectivity is not a necessity in every type of network. This has led to a totally progressive kind of social network called mobile social networks (MSNs). MSNs can be viewed as modern kinds of delay-tolerant networks (DTNs) in which mobile users interact with each other to share user-centric data objects among interested observers.

MSNs have mainly been introduced by combining social networks from social science and mobile communication networks. In this way, not only can users utilize the knowledge of their relationship to improve the efficiency and effectiveness of network services, but they can also access, share, and distribute data in a mobile environment by exploiting the social relations. Compared to online social networks, MSNs reflect social interactions more realistically. This is because mobile phones are more common than Internet-based communications in everyday life.

There is a consensus among researchers who discern two groups of MSNs [1], [2]: *centralized* MSNs, which use social network services for acquiring information through mobile devices, and *decentralized* MSNs in which communication is performed opportunistically using wireless technologies such as Bluetooth or Wi-Fi. Decentralized social networks have a lot in common with opportunistic networks (OppNets) [3] and DTNs [4], [5]. However, in contrast to OppNets and DTNs, decentralized networks mostly attempt to exploit complex social similarities and behaviors (such as community and friendship) between mobile carriers to establish social-aware protocols and approaches and enhance data-forwarding models.

As MSNs inherit part of their characteristics from OppNets and DTNs, a quick overview of safety challenges and solutions in OppNets and DTNs [6]–[9] is necessary. However, having included social aspects, MSNs encompass more complex and correlated challenging safety problems than other types of networks, although like any other modern model of technology, MSNs depend upon time to be totally safe and immune. To overcome such a goal, a clear classification on safety issues in MSNs seems to be a necessity. It makes it possible to identify concerns, investigate brand new solutions, and evolve suitable approaches in order to prevent any obstacle from reflecting on major parts of safety. Safety problems involve a variety of closely related aspects. In this paper, we will focus on three of them, i.e., trust, security, and privacy, due to their increasing importance in the field of MSNs.

Safety can be described as the condition of being protected against different types of failure, damage, error, accidents, harm, or any other nondesirable event. In early works, like [6], the term security was used to convey this meaning, but later, with the emergence of various networks and safety issues, security has been specified to a more technical concept. What we mean by safety in this paper is to be in control of recognized hazards and to achieve an acceptable level of trust, security, and privacy. This can take the form of protection from an event or exposure to something that causes damage. It includes the

Manuscript received November 22, 2012; revised April 23, 2013 and September 5, 2013; accepted September 18, 2013. This work was supported in part by the Natural Science Foundation of China under Grant 60903153, by the Liaoning Provincial Natural Science Foundation of China under Grant 201202032, and by the Fundamental Research Funds for the Central Universities under Grant DUT12JR10.
Y. Najaflou, B. Jedari, and F. Xia are with the School of Software, Dalian University of Technology, Dalian 116620, China (e-mail: yasharnajafloo@yahoo.com; bjedari@mail.dlut.edu.cn; f.xia@ieee.org).
L. T. Yang is with the School of Computer Science and Technology, Huazhong University of Science and Technology, Wuhan 430074, China, and also with the Department of Computer Science, St. Francis Xavier University, Antigonish, NS B2G 2W5, Canada (e-mail: ltyang@gmail.com).
M. S. Obaidat is with the Department of Computer Science and Software Engineering, Monmouth University, West Long Branch, NJ 07764 USA (e-mail: obaidat@monmouth.edu).
Color versions of one or more of the figures in this paper are available online at http://ieeexplore.ieee.org.
Digital Object Identifier 10.1109/JSYST.2013.2284696

protection of users or their possessions (information, identity, location, etc.).

Trust in MSNs is defined as the willingness of a particular node to rely on the actions of a node or nodes in a mobile social environment. This reliance directly impacts the network safety, and this is the main reason that we consider trust as the first component of safety in MSNs. Being specific cases of peer to peer (P2P) networks, MSNs suffer from the lack of designated routers. This lack of infrastructure turns MSNs to a vulnerable target for various attacks due to the existence of malicious nodes which do not comply with the network protocol. This can directly endanger trustworthiness. Although specific protocols have been designed to achieve malignity prevention [10]–[22], different models have been integrated to develop the assorted trustworthiness [23]–[29] throughout the mobile networks. The cooperation between nodes is therefore vital for independent communication from which forwarding strategies are adopted. However, the increase of selfish behavior among nodes, as a result of the scarcity of resources, can endanger this cooperation in MSNs. In order to overcome such an obstacle, selfishness discouragement [30]–[37] and cooperation enforcement [38]–[40] have been designed and applied in the network to improve confidence among nodes so that nodes are able to rely on one another and perform networking operations efficiently.

Security is another component of safety which is not unique to MSNs. The fundamental aim of security is to protect the information and the resources against attacks and misbehavior throughout the network. In order to unify endpoint security technology and define security policies, scholars have equipped MSNs with access controls [41]–[47] as a means of classical defense. However, one cannot suffice to this method alone. Other precautions such as confidentiality techniques [48]–[55] are needed to prevent information disclosure and manipulation of unauthorized individuals or groups. Despite the implementation of both classical defense and confidentiality schemes, there are always some leaky links that allow an intruder to gain unauthorized access to a protected network and exploit it. To tackle such attacks, a set of intrusion detection systems (IDSs) [56]–[58] has been used to monitor a network and produce reports about any malicious activities or policy violations throughout the network.

Privacy is the last aspect of safety in MSNs that has recently gained unprecedented attention. This is particularly because of the social aspect of MSNs in which information relevance like users' location and identity is considered critical issues for both the attackers and system administrators. Many researchers have suggested ways to reveal users' information selectively and for them to remain unnoticed or unidentified over the network. To do this, privacy preferences are generally specified to obfuscate [59]–[66] users' private information and present it in a coarser and falsified manner. Moreover, fairness encouragement [67]–[71] strategies have been included to prevent heavy congestion in a particular collaborative node. The goal is to encourage nodes to forward messages and distribute their private information equally. This prevents malicious nodes from intruding into the network and gaining an unauthorized access to valuable resources. Furthermore, as attackers are in direct correlation with the personal profiles in MSNs, methods for private matching [72]–[82] have been designed to let two users conceal their personal profiles while in connection. To deliver location-based services in MSNs, privacy should be maintained. It can be achieved through different techniques such as obfuscation-based schemes [83]–[85], social-based schemes [86]–[89], dynamic pseudonymity [90]–[93], and key anonymity [94]–[96]. Meanwhile, the information of each individual should be protected while communicating with the other, so communication privacy [97], [98] should be considered vital for every network.

Due to the epic correlations between safety issues and the inordinate area that they cover, a clear categorization to distinguish their characteristics has seemed too ambitious for researchers. This paper aims to provide a clear categorization on safety challenges and a deep exploration over some recent security, privacy, and trust solutions in MSNs. It narrows the safety challenges and techniques down from OppNets and DTNs to MSNs and tries to cover all the work proposed around security, privacy, and trust in MSNs. To the best of our knowledge, it is the first attempt to classify safety challenges in MSNs.

The remainder of this paper is organized as follows. Section II offers an overview of safety challenges and solutions in MSNs. Section III introduces trust as the first element of safety in MSNs and focuses on its characteristics, challenges, and given solutions. An ideal categorization of all the different forms of trust is presented in this section. Section IV addresses security as the second key element of safety issues in MSNs. In this section, security-preserving approaches are discussed in detail, and a clear classification of related works is displayed. Section V concentrates on privacy as the third main category of safety challenges and solution in MSNs. To conclude, several major open research issues are discussed, and future research directions are outlined in Section VI.

## II. Safety Challenges in MSNs

MSNs are considered as a particular type of OppNets, and they share a lot of common characteristics with OppNets and DTNs. As a result, MSNs cover some of the safety concerns related to OppNets as their challenges are partially the same. The first set of works to distinguish safety challenges in an OppNet goes back to the introduction of OppNets. Lilien *et al.* [6] were the first to propose OppNets along with a classification of safety challenges, containing privacy and security, in six different steps. They proposed a general safety scheme for OppNets in five mandatory functions in the absence of initial authentication mechanism. Another categorization was deployed in [7] where safety challenges in opportunistic people-centric sensing were studied and general suggestions to make solutions were discussed. Further attempts to classify safety issues in opportunistic communication were made by Shikfa [8]. This work itemizes basic challenges in OppNets into authenticity, confidentiality, cooperation enforcement, trust establishment, and integrity privacy according to the concerns provided.

MSNs include more vital and complex safety concerns in comparison to other resembling networks and contain tons of safety challenging problems, specifically in trust, security, and privacy. There have been few attempts proposing a clear categorization of safety in MSNs. To take an example of these efforts, Beach *et al.* [9] presented issues around privacy and security for MSNs, along with some methods and implementation for their solutions. They classified problematic issues in three groups, namely: direct anonymity issues, indirect or k-anonymity issues, and attacks (eavesdropping, spoofing, replay, and wormhole). Furthermore, they expanded their proposition by designing an identity server (IS) which adopts established privacy and security technologies to provide solutions for these problems.

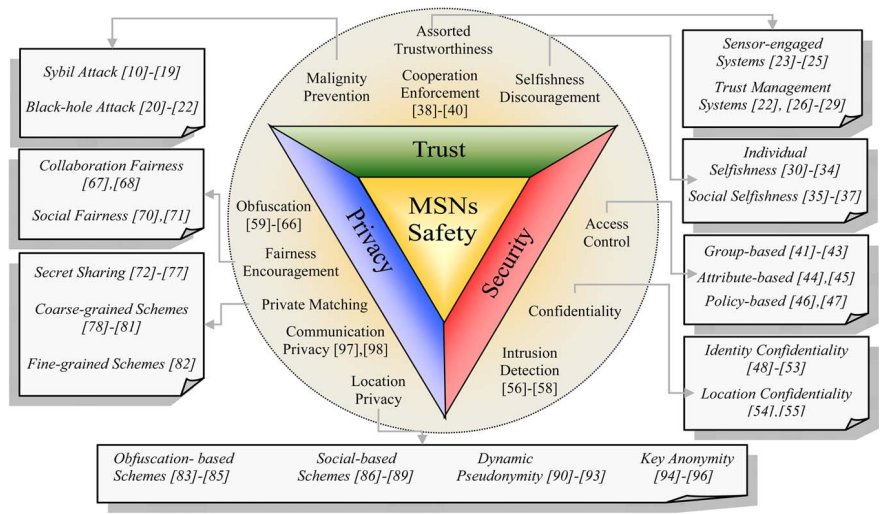

Fig. 1. Safety challenges in MSNs in three main classes, namely, trust, security, and privacy, expanded in ten subclasses and their belonging attributes to provide a safe mobile social environment for users' communication.

Although some efforts have been made to take safety issues into consideration and make a classifiable observation to enumerate safety challenges and solutions in MSNs, they have been neither comprehensive nor detailed. As far as we know, there has never been an obvious categorization followed by a comprehensive clarification on MSN safety issues. To do this, we classify these issues in three main groups, *trust*, *security*, and *privacy*, and explain noble and novel approaches for possible solutions, as outlined in Fig. 1.

## III. TRUST

Trust is a critical determinant of sharing information and developing new relationships in a network. In other words, trust is based on the reputation between individuals and is a capital asset that people may invest great amount of resources in building and that is acquired slowly but can be destroyed very quickly. In traditional networks, trust relies mainly on the infrastructure, and this infrastructure is trusted by end users to fulfill the routing task and provides naming service which also simplifies the establishment of trust between users. Furthermore, when higher level of trust is required, the network infrastructure can be complemented by a safe infrastructure. In OppNets, there is no routing infrastructure with dedicated routers, and peer nodes act as message carriers and forwarders instead. Not only is there no naming service accessible, but also identifiers are meaningless from a trust perspective. It is therefore important in OppNets to first build trust among parties so that they can rely on one another. Including social context into the information generation and delivery process provides greater assurance for receiving trustworthy content. Maintaining trust demands special mechanisms and protocols in MSNs. We categorize social trust schemes according to their mechanisms for the following: 1) *malignity prevention*; 2) *assorted trustworthiness*; 3) *selfishness discouragement*; and 4) *cooperation enforcement*.

### A. Malignity Prevention

Trust retention over interactive networks partially relies on the robustness of protocols to defend against malicious nodes. These nodes give erroneous responses to a query, either by returning false data or returning false routing information. Among different kinds of attacks, there are two common attacks in MSNs: Sybil attack and black-hole attack. The following section discusses the mechanisms of these two types of attacks along with some solutions proposed to resist them.

*1) Sybil Attack:* Sybil attack [10] is one of the basic malicious actions in MSNs which describes an attempt to create many identities in order to gain a larger effect in a reputation system (RS). It is very common to employ a trusted certification center to verify the physical identity for preventing multiple-identity attacks. These types of attacks usually use a single malicious node to confuse neighboring nodes and equip a malicious user with the capability to create a relationship with honest users. When one honest user is compromised, a malicious user will gain special privileges which can be used for an attack. Fig. 2 demonstrates an example of this attack in a mobile social environment.

In order to detect such attacks, Piro *et al.* [11] observed that Sybil users should only communicate serially as it can cause much fewer collisions at the media access control layer. Later, Capkun *et al.* [12] made users able to build certificate chains similar to the pretty-good-privacy algorithm based on the assumption of unconditional transitivity of trust along the chain paths.

Another significant work to detect Sybil attacks was made in [13]. The main goal of this work is to assess a genuine user in MSNs with honest intentions by limiting the maximum number of Sybils independently. In addition, it encompasses two distinguished complementary appendages: the explicit and the implicit social trust. Explicit social trust is based on consciously established friend ties and calculates trust as a function of hop distance and interconnection. It conveys trust which, in terms of identity, is not Sybil and verifies the honesty of the user's intentions. Implicit social trust leverages mobility properties to convey trust in the originality of identities due to their persistency. Their further proposition was to optimize the trust level of the initial version of PodNet [14]. In this version, content publication was done anonymously, making it an ideal platform for spams and illegal contents to be spread. They proposed an integrated safe framework called PodNetSec [15] which contains three types of channels, namely, open, restricted, and closed channels, as shown in Fig. 3. Closed channels allow the private and encrypted dissemination of content in a limited group. Restricted channels only allow authorized users to publish content but allow everybody to consume it. Open channels allow every user to consume as well as create new content. Although this work mainly focuses on trust in OppNets, having the ability to maintain trust in an anonymous content-publication environment makes it possible to be implemented in MSNs.

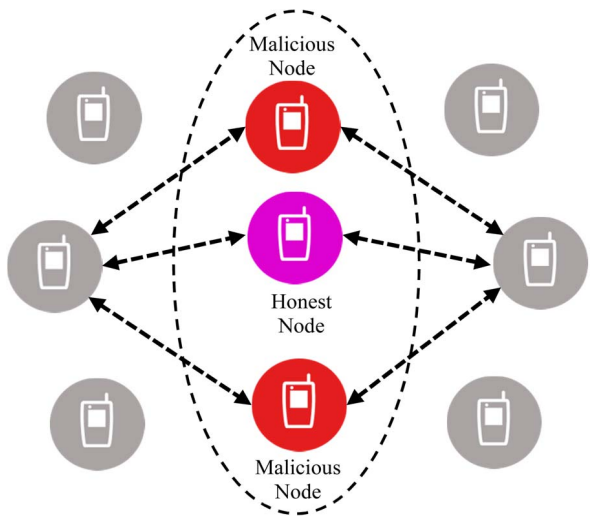

Fig. 2. Example of Sybil attack.

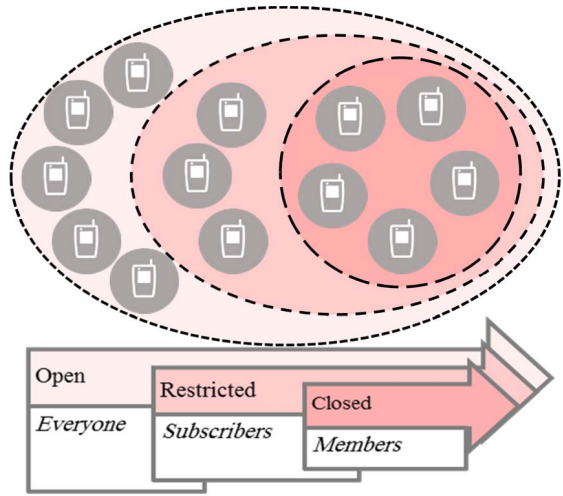

Fig. 3. Interconnection of safety channels in PodNetSec [15]: Open, restricted, and closed channels.

Although malicious users can create many identities in a Sybil attack, they can only establish few trust relationships. In other words, there is always a small cut in the graph between the Sybil nodes and the honest nodes. SybilGuard [16] exploits this property to limit the number of identities that a malicious user can create. However, this scheme has been designed for P2P networks and is not applicable in mobile networks. MobID [17] is a decentralized defense for portable devices which differentiates between friends' network (containing honest nodes) and foes' network (containing suspicious nodes). By reasoning on these two networks, the node is able to determine whether an unknown node is carrying out a Sybil attack. Another important approach to defend against Sybil attacks in social networks is called SybilDefender [18]. It can identify the Sybil nodes and detect the Sybil community around a Sybil node, even when the number of Sybil nodes introduced by each attack edge is close to the theoretically detectable lower bound. Based on performing a limited number of random walks within the social graphs, SybilDefender is efficient and scalable to large social networks.

Most of social network-based Sybil defense mechanisms exploit the algorithmic properties of social graphs to infer the extent to which an arbitrary node in such a graph should be trusted. However, these approaches neither consider the different amounts of trust represented by different graphs nor consider the inherent trust properties of the graphs that they use. To tackle this weakness, Mohaisen *et al.* [19] evaluated the performances of anti-Sybil algorithms and compared them using the cost function. The cost function is the required length of random walks in the social graph to accept all honest nodes. They found that the cost function in high-trust graphs is much higher than that in low-trust ones in a social network. They proposed several designs to model trust in social networks to increase the cost function of social graphs with low-trust value which directly decreases the advantage of the attacker.

*2) Black-Hole Attack:* Black hole [20] is a type of denial of service (DoS) attack which targets trust. In this type of attack, a malicious node, in reply to its given path request (PREQ), sends a forged path replay packet to a source node that initiates the route discovery. This is because the malicious node pretends to be a destination node itself or an immediate neighbor of the destination node. As a result, the source node will forward all of its data packets to the malicious node. Fig. 4 describes an example of this attack.

To prevent such an attack, Li and Das [21] designed a trust-based framework and later integrated it [22] with a large family of existing single-copy data-forwarding protocols to provide a comprehensive evaluation of an encounter's ability to deliver data. Their framework not only resists black-hole attack but also contracts arbitrarily forwarding attacks, effectively contributing to three major components: positive forwarding message, a trust-based framework, and data-forwarding protocol.

### B. Assorted Trustworthiness

Assorted trustworthiness refers to a system or group of systems that aim to maintain trust across a network by integrating different models, schemes, or mechanisms. In the following two sections, we will discuss these systems in two different

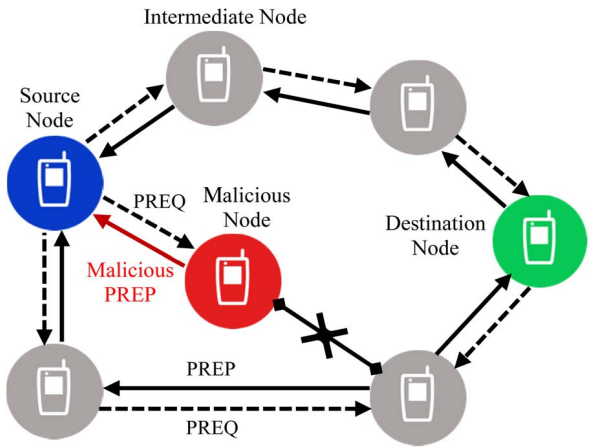

Fig. 4. Example demonstrating a black-hole attack in which source node is attacked in reply to its path request.

groups, including *sensor-engaged systems* and *trust-management systems*.

*1) Sensor-Engaged Systems:* While sensor-engaged applications do not seem to be widely deployed opportunistically, there have been efforts in creating opportunistic sensing systems. An interesting project is CenceMe [23] which facilitates sharing information throughout the social network. The main strategy is to leverage the growing integration of sensors into off-the-shelf consumer devices. Later, Zhao *et al.* [24] presented a collaboration model that can further be developed and used in opportunistic sensing systems to obtain a neutrality policy from the injecting sensors or users.

To retrieve social informatics properly and safely, it is important to design a system that enhances the management of social data. To do this, a social computing paradigm was proposed [25] which is a concept to promote innovative and collaborative cybersecurity models, particularly pervasive trustworthy computing in opportunistic sensing networks. It takes advantage of the integration of sensors, applications, and online social networks. Given the availability of both digitized social informatics and sensor contents, this approach allows users to examine these sources simultaneously.

*2) Trust-Management Systems:* The aim of trust management is to aid the automated decision-making process with results of trust assessment in social networks. One of the important studies in this area is a framework for OppNets called trust-management system [22] which basically consists of two elements, a watchdog and an RS. The watchdog monitors the real routing behavior of a node, feeding that information into the RS to update the reputation of that node. The RS updates reputation opinions based on the direct observation of the watchdog. In addition, it integrates reputation by combining the indirect information, from other members, with first-hand information. Another function of RS is to perform reputation aging. The aging is a self-updating mechanism and is used when fresh first-hand information is not available for a long period of time.

Although many trust-management systems have been proposed, a few of them can be applied to MSNs due to their unique communication characteristics. To achieve this goal, a trust-management system called MobiTrust [26] was proposed which establishes secure, reliable, and accurate trust relationships between network participants in MSNs. It encompasses three key factors associated with the similarity of user profile, reputation, and history of friendship.

Managing reputation in large-scale trust-management systems can be problematic. This is especially true because computing the marginal probability functions in large-scale systems is computationally prohibitive. To address this problem, Belief Propagation-Iterative Trust and Reputation Management (BP-ITRM) [27] was proposed. It is an iterative probabilistic and belief propagation-based approach in the design and evaluation of trust and reputation management systems. It utilizes the belief propagation algorithm to efficiently compute these marginal probability distributions. This approach models the RS on a factor graph, which chiefly has two benefits. The first advantage of using such a graph is the ability to obtain a qualitative representation of how the consumers and service providers are related on a graphical structure. The second advantage is that, by using such a factor graph, the global functions can be included in products with simpler local functions, each of which depends on a subset of the variables. In addition, BP-ITRM allows a message to be passed between nodes in the graph in order to compute the marginal probability distribution functions of the variables representing the reputation values. Furthermore, BP-ITRM is reliable in filtering out malicious reports, and it reduces the error in the reputation values of service providers caused by the high probability of malicious raters.

Many trust-management models try to minimize the threats to evaluate the trust and reputation of the interacting agents. However, when malicious agents start to behave in an unpredictable way, they fail to properly evaluate trust. Moreover, most of these models are not capable of distributing workload among service providers. To surmount these problems, a dynamic trust computation model called SecuredTrust [28] was suggested. It is a comprehensive quantitative model for measuring such trust with a load-balancing algorithm based on different factors to perform workload distribution effectively.

A perfect example of trust-management systems in service-oriented MSNs is Trustworthy Service Evaluation (TSE) [29] in which service providers can collect and store users' reviews about their services without requiring any third trusted authority. The service reviews can then be made available to interested users for making service selection decisions. TSE consists of two subsystems: basic TSE and Sybil-resisted TSE. Basic TSE enables users to submit their reviews by using hierarchical and aggregate signature techniques in a distributed and cooperative manner. It restricts the service providers by not allowing them to reject, modify, or delete the reviews. Sybil-resisted TSE enables the detection of typical Sybil attacks. It reveals the real identity of the user who intends to generate multiple reviews toward a vendor in a predefined time slot with different pseudonyms.

*C. Selfishness Discouragement*

While intermediate nodes are expected to help each other to store, carry, and forward packets in a cooperative and opportunistic fashion, in reality, these nodes are so rational that they communicate with relative nodes rather than devote their own valuable resources to serve the network. This phenomenon is called *selfishness* and has been classified into two basic groups, namely, *individual selfishness* and *social selfishness*. We will elaborate these types of selfishness and some recent approaches to handle them in the following two sections.

*1) Individual Selfishness:* Individual selfish nodes tend to get help from others rather than helping them to accrue their own orientations. To impose cooperation among individual selfish nodes, different incentive schemes were proposed. These schemes mainly try to let the nodes gain utility according to the degree of help that they spread over the network. Reputation-based schemes [30] make individual nodes responsible for monitoring traffic and reputation in the vicinity. Their aim is to pinpoint uncooperative nodes and exclude them from the networks. On the other hand, credit-based schemes [31] regulate the packet forwarding relationships among different nodes using virtual currency.

An interesting example on reputation-based schemes is SUCCESS [32]. It is a secure user-centric and social-aware reputation-based incentive scheme for DTNs that allows a node to manage its reputation evidence and demonstrate its reputation whenever necessary. It mainly relies on neighboring nodes to monitor the traffic and keep track of each other's reputation. In this paper, two concepts, notably self-*check* and *community check*, were defined to evaluate reputation and speed up the reputation establishment. Probably, the best example of a credit-based scheme is SMART [33]. It uses credits, composed of multiple layers, to provide incentives to selfish nodes. Credits are distributed among DTN nodes through a bundle-forwarding cooperative manner without dependence on any tamper-proof hardware. Using its concentration technique, SMART makes each intermediate node able to generate a new credit layer, based on the previous layers. This credit generation is completed by adding a nonforgeable digital signature which implies that the forwarding node agrees to provide forwarding service.

However, none of the reputation-based and credit-based schemes is specifically designed to solve individual selfish behavior in MSNs. Give2Get [34] is probably the only individual selfishness solution introduced specifically for MSNs which contains two forwarding protocols to force faithful behavior. It considerably reduces the number of forwarding messages, which leads to positive side effects and performance improvements.

*2) Social Selfishness:* Social selfishness is closely related to not only the willingness intention for forwarding but also the trusted relationship between nodes. Social selfish nodes prefer to ask the influential nodes with more social relations to help them forward the packets. This type of selfishness is widely evident in MSNs and causes serious moderations in the network performance. One of the first approaches in this area was proposed in [35] which studies the impact of different distributions of selfishness, including the impact of topologies and traffic patterns. This work evaluates the system performance using four experimental human mobility traces with uniform and community-biased traffic patterns. The authors found that, due to the nature of multiple paths, MSNs are resilient against the distributions of social selfish behavior.

In order to resolve such selfishness, trust-based strategies often establish trusted relationship to complete trusted routing by coping with social selfishness. One of these approaches is a social selfishness aware routing called SSAR [36]. It allows user selfishness by considering users' willingness and their contact opportunity to prepare a better forwarding strategy than purely contact-based approaches. Later, Xin *et al.* [37] introduced a virtual bank to avoid social selfishness. It uses a reward system and distributes it throughout the participating nodes in the packet forwarding process. Taxation strategy is also adopted to redistribute rewards among poverty isolated nodes that have the potential to become the malicious nodes. This protects consequential internal threats caused by the social discrimination.

*D. Cooperation Enforcement*

Cooperation is the process of working or acting together which involves things working in harmony and can be accomplished by handling shared time and resources simultaneously. The fundamental challenge in mobile networks, regardless of the application, is to enable node cooperation to forward a message.

One of the finest mechanisms for node cooperation was proposed in [38]. The main aim of this work is to investigate the potential impact of the lack of trust on node cooperation. It leverages social information and proposes six trust-based filters to establish trustworthy communications over mobile OppNets. The six filters couple three socially aware estimators of trust, including common interests, common friends, and the distance in the social graph, with two major techniques of trust establishment, including relay-to-relay and source-to-relay. It has been shown that the trust filters yield a fair tradeoff between trust and success rate.

In order to establish cooperation between nodes in OppNets, the authors in [39] highlighted trusted devices through the direct and the indirect observation and created policies in which nodes compromise to obtain a service. In other words, their method obtains direct past experience using direct observation while indirect ones are for using recommendations. A noncollaborative fading parameter that decreases the reputation values of users is considered. Another strategy to make users collaborate, which seems applicable for MSNs, was proposed in [40] in which users are obliged to be cooperative by handling essential and effective messages throughout the network while messages are either primary or secondary. Messages that are considered to be essential for a device itself are called primary, whereas secondary messages are useful for other users and carrying them is proof of the cooperation between the devices. By using a barter exchange, a user receives a message only if it provides the same number of messages to the other participant.

In Table I, a demonstration of our categorization in trust-based schemes is given. We mostly focused on the functionality of socially compatible schemes rather than some schemes that target trust in an unsocial manner.

IV. Security

Network information security is concerned with the confidentiality, integrity, and availability of data, regardless of the form that the data may take. However, the security services of

TABLE I
DEMONSTRATION OF SOME RECENT SAFETY SCHEMES IN TRUST CATEGORY IN MSNs

| Scheme | Functionality | Characteristics ||||||| 
|---|---|---|---|---|---|---|---|---|
| | | malignity prevention || assorted trust || selfishness discourage || cooperation enforcement |
| | | Sybil attack | black hole attack | Sensor systems | trust management | individual | social | |
| Social Trust [13] | Confines the maximum number of Sybil independently using explicit and implicit trust. | √ | × | × | × | × | × | × |
| PodNetSec [15] | Leverages trust layers in three channels by maintaining different level of restrictions for publishers. | √ | × | × | × | × | × | × |
| MobID [17] | Puts honest and suspicious nodes in two separate networks so nodes can determine whether an unknown node is carrying out a Sybil attack. | √ | × | × | × | × | × | × |
| SybilDefender [18] | Detects Sybil nodes and the community around them by leveraging network topologies. | √ | × | × | × | × | × | × |
| RADON [21] | Monitors the forwarding behavior of a node using the number of times of previous encounters as the metric to select the next qualified forwarder. | × | √ | × | × | × | × | × |
| Data Forward. FRMW [22] | Blocks black-hole attacks by contributing positive forwarding message, a trust-based framework, and data forwarding protocol. | × | √ | × | √ | × | × | × |
| CenceMe [23] | Leverages integration of sensors into off-the-shelf consumer devices. | × | × | √ | × | × | × | × |
| Socialwiki [24] | Obtains a neutrality policy from the injecting sensors or users. | × | × | √ | × | × | × | × |
| Trust Opp Sensing [25] | Cultivates online social pervasive trustworthy computing by integrating sensors and applications. | × | × | √ | × | × | × | × |
| MobiTrust [26] | Establishes reliable and accurate trust using profile similarity, reputation, friendship history. | × | × | × | √ | × | × | × |
| BP-ITRM [27] | Computes marginal probability distributions by using a belief propagation algorithm and modeling a reputation system on a factor graph. | × | × | × | √ | × | × | × |
| SecuredTrust [28] | Performs workload distribution by measuring trust with a load-balancing algorithm. | × | × | × | √ | × | × | × |
| TSE [29] | Makes service providers collect and store users' reviews about their services without requiring any third trusted authority. | × | × | × | √ | × | × | × |
| SUCCESS [32] | Enables all nodes to manage their own reputation and gives neighboring nodes the ability to monitor traffic by keeping track of each other's reputation. | × | × | × | × | √ | × | × |
| SMART [33] | Allows credits to be distributed among nodes through a bundle forwarding cooperative manner without reliance on any tamper proof hardware. | × | × | × | × | √ | × | × |
| G2G [34] | Forces faithful behavior to leave positive side-effect to improve performance by reducing message ratios. | × | × | × | × | √ | × | × |
| SAMS [35] | Evaluates system performance using experimental human mobility traces with uniform and community-biased traffic patterns. | × | × | × | × | × | √ | × |
| SSAR [36] | Allows user selfishness, considering users' willingness and their contact opportunity. | × | × | × | × | × | √ | × |
| Internal Threat Avoidance [37] | Distributes rewards among packet forwarders, using virtual bank, and adopts taxation strategy to redistribute rewards among isolated nodes. | × | × | × | × | × | √ | × |
| Social-based Trust [38] | Investigates the potential impact of lack of trust on node cooperation and leverages social information through six trust based filters. | × | × | × | × | × | √ | √ |
| Informal AssessRisk [39] | Creates policies so nodes can compromise to obtain a service using past experience or indirect recommendations. | × | × | × | × | × | × | √ |
| Barter Trade Msg. Delv. [40] | Obliges cooperation using barter exchange in which a user receives a message only if it provides the same number of messages to the others. | × | × | × | × | × | × | √ |

"√" – if the scheme satisfies the property, "×" if not

OppNets are not altogether different from those of other network communication paradigms. The general goal of security maintenance is to protect the information and the resources from attacks and misbehavior throughout the network. Requirements to ensure an effective security paradigm can be explained as availability, authenticity, confidentiality, and integrity. On the other hand, MSNs are defined as specific opportunistic P2P networks which do not have an access point in between them. Having little protection, these types of networks usually employ encryption methods to provide security. An attacker, however, may try to break the encryption of the network; hence, security policies were defined. These policies address constraints on functions and access of external attacks, adversaries, and unauthorized users. To offer a comprehensive study on MSN security issues, we put them in three groups: 1) *access control* to exert control over interactive nodes; 2) *confidentiality* to ensure that a given message cannot be understood by anyone else other than its desired recipients; and 3) *intrusion detection* to monitor the network for malicious activities or policy violations. In the following sections, we will elaborate on these three categories.

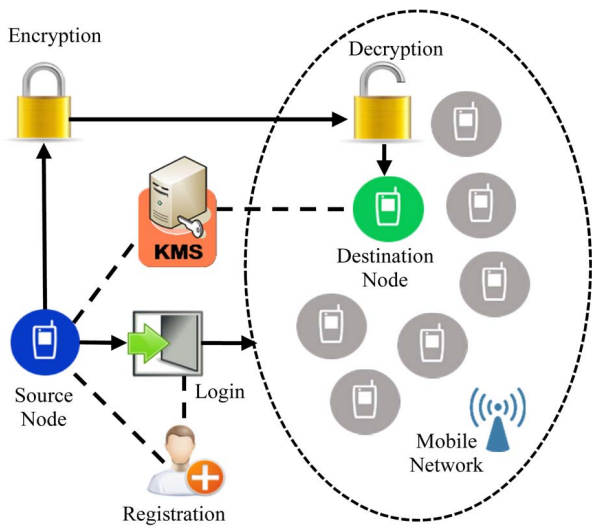

Fig. 5. Example of access control establishment to secure a mobile network using authentication and key management system.

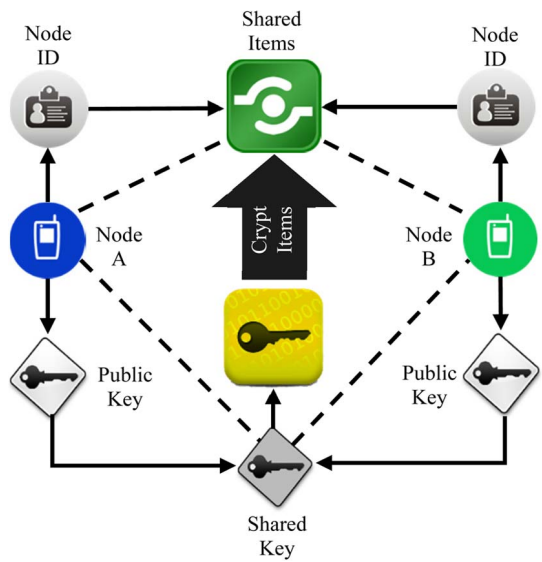

Fig. 6. Group-based access control using fine-grained access policy [40] to share items securely in a mobile network.

## A. Access Control

Network access control is an approach to unify endpoint security technology, user or system authentication, and network security enforcement. It uses a set of protocols to define and implement policies that describe how to secure access to network nodes by devices when they initially attempt to access the network. However, using it in a mobile deployment involves challenges. When a user is denied access because of a security concern, the productive use of the device is lost, which can affect the ability to complete a job or serve a customer. Modern mobile network access control solutions give system administrators greater control over when, how, and whether to remediate the security concern. They also combine access policies with cryptographic approaches and key management techniques to maintain an efficient level of security over the network. Fig. 5 illustrates an example of the collaboration of authentication and key management system to establish a safe and secure communication over a mobile network. In general, access control-related approaches in MSNs fall into three categories, depending on their mechanisms: *group-based access control, attribute-based access control*, and *policy-based access control*. The next three sections discuss these categories in detail.

*1) Group-Based Access Control:* The base of group-based access control in MSNs lies in a general cryptographic scheme for OppNets called HiBOp [41] which divides users into communities in order to communicate safely. It uses public-key cryptography and selects specific nodes to forward messages between communities. This approach provides basic security for an opportunistic environment. It is therefore considered an important work because it was the base of many other consequent methods. Group-based cryptography is one of these methods by Shikfa *et al.* [42] which is first used for OppNets. The aim of this work is to prevent data from being accessed by different groups, using multiple levels of cryptography. Later, Graffi *et al.* [43] introduced a solution to enable authenticated and secure communication between MSN users. This method establishes a trust infrastructure which provides personalized fine-grained data access control. Using this method, all communications are encrypted with the public key of the receiver, whereas secure and authenticated communications are provided once the node identification (ID) of the receiver is known. Any node may retrieve and replicate these data, but only privileged users can decrypt it. Fig. 6 explains this approach, where two nodes share items and implement fine-grained access policies and group-based encryption to maintain a secure connection.

*2) Attribute-Based Access Control:* This kind of access control uses attribute-based encryption techniques in which a sender encrypts a data packet with an access policy and a receiver decrypts the packet and reads its content only if its attributes satisfy the access policy. An example of the schemes which use such techniques is called secure symptom-based handshake (SSH) [44]. It is a privacy-preserving personal profile matching scheme which tackles challenging security issues for m-Healthcare social network. Using this scheme, each patient is granted with a pseudo-ID and its own private key corresponding to his/her symptom. When two patients meet, they can use their private keys to make mutual authentication, only if they have the same symptom. SSH satisfies three properties, namely, correctness, impersonator resistance, and detector resistance.

Although SSH is very efficient in a mobile social environment, several security and privacy issues (i.e., resisting inside attacks) must be addressed through MSNs in advance. To do this, Xiaohui *et al.* proposed an efficient and secure user revocation scheme [45]. It uses attribute-based encryption techniques to enable a trusted authority in order to monitor the data decryption capability of mobile social users. This scheme not only does disable malicious users from decrypting any data packet but also encourages proper user behavior to decimate inside attacks. It has also the ability to resist attribute collusion and revoke collusion attacks.

*3) Policy-Based Access Control:* Policy-based access control provides a powerful and flexible means of protecting data, down to the row level. Using this type of access control, administrators can define security policies that are based on the value of individual data elements, and the server can enforce these policies transparently. In other words, it simplifies both the security administration of an adaptive server installation and the application development process because it is the server that enforces security.

An example of policy-based access control for P2P networks was proposed in [46]. It is a framework for dynamic multilevel access control policies based on trust and reputation. This framework allows a group to switch between policies over time, influenced by the system's state or environment. Based on the behavior and trust level of peers in the group and the current group composition, it is possible for peers to collaboratively modify policies such as join, update, and job allocation.

In MSNs, Hachem *et al.* [47] propose a policy framework for controlling access to social data in mobile applications. This framework allows the representation of expressive policies based on users' social interactions while keeping policy model compatibility. In other words, it combines expressivity with personalization in MSNs. They integrate the policy framework as part of a middleware for managing mobile users' social ecosystems to enforce policy-based access control on heterogeneous devices with minimal deployment effort.

### B. Confidentiality

Confidentiality is the term used to prevent the disclosure of information to unauthorized individuals or systems. In other words, it means that no one can gain, read, or manipulate information other than for whom it is intended. Basically, confidentiality is achieved in two steps: encryption and decryption. Using encryption, the sender converts plaintext to ciphertext with the aim of rendering it unintelligible to parties except the intended recipient. Using decryption, ciphertext is rendered intelligible to the intended recipient by converting it back to the plaintext. The foundations of confidentiality have not been changed in OppNets, DTNs, and MSNs. Therefore, owing to the vast utilization of asymmetric cryptography in this area, new categorizations have been made. The next two sections highlight our confidentiality in two basic groups: *identity confidentiality* and *location confidentiality*.

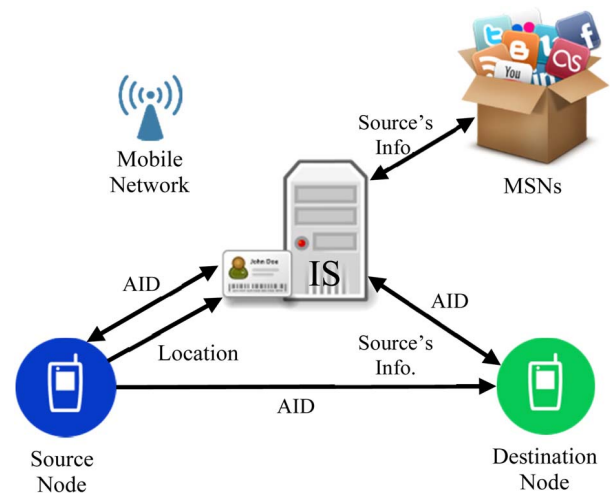

Fig. 7. Use of an IS, proposed in [6], to establish identity confidentiality in MSNs.

*1) Identity Confidentiality:* The identity of the nodes in a mobile network is an important item which should be protected. To do this, many approaches have been made. To take an example, Spring [48] is a privacy-preserving protocol for vehicular DTNs which aims to achieve conditional privacy preservation and resist most of the attacks. Another example is the work proposed by El Defrawy *et al.* [49]. This work focuses on the problem of initial secure context establishment in DTNs and allows users to leverage social contacts to exchange confidential and authentic messages. To develop security infrastructure in DTNs, Patra *et al.* [50] suggested the employment of identity-based cryptography to let a source derive the destination public key from some associated identity strings. A complementary work in this area is a dynamic virtual digraph model [51] for public-key distribution study. Extending the graph theory and distinguishing between owners and carriers, this model presents a public-key distribution scheme for pocket DTNs based on two-channel cryptography.

Identity confidentiality establishment in unstructured Opp-Nets, like MSNs, is discussed in a project called *cloud* [52] which is an end-to-end transportation standard. This project stresses ambiguity and intelligence resistant search functionality for the semantic OppNets. Ambiguity is satisfied in the emergence of high connectivity among the same hosts. It results in hosts hiding their identity within groups of semantically close network. A cryptographic standard also takes care of counterintelligence by protecting unidentified transmission between the source and the destination. The mechanism used in this project is independent of any earlier sent messages. This is can be very beneficial because it guarantees the confidentiality of the supplies.

Being independent, however, can have its own drawback. This is because it avoids centralized communication and assures competence to break strong transmission between hosts. To compensate this flaw, Beach *et al.* [9] proposed the use of an anonymous identifier (AID). Generally, AID is a nonce that is generated by a trusted server, called the IS, to provide solutions for clear text exchange threats. As depicted in Fig. 7, before a source mobile node advertises the user's presence to other nearby nodes and stationary devices, it securely contacts the IS to obtain the AID. The IS generates a new AID, using a cryptographic hash function, and associates it with the destination mobile node that made the request. Then, the destination node or stationary device discovers this AID sharing service on the source node. Eventually, it establishes a connection to obtain the shared AID, so the source node would be able to make a request for another AID from the IS, which is to be shared with the next mobile or stationary device. This scheme was later expanded by suggesting a secure social-aware [53] framework to make interactions of real-world location-based services of MSNs. This framework helps exchange an encrypted nonce combined to a verified user location without any use for user privacy and security adjustment.

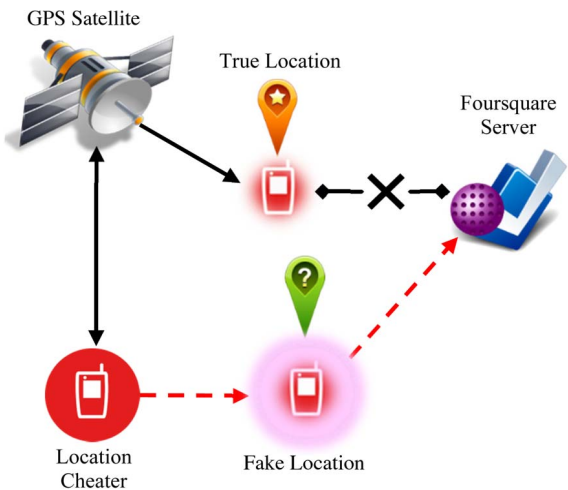

Fig. 8. Example of location cheating against a foursquare server.

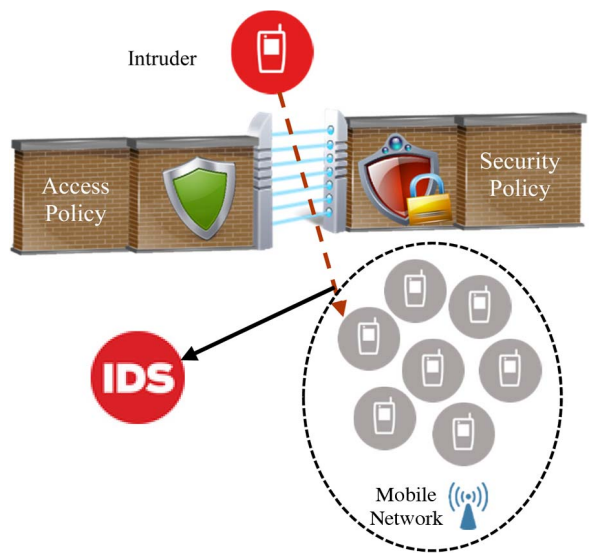

Fig. 9. Example of bypassing security and access by an intruder caught by IDS in a mobile network environment.

*2) Location Confidentiality:* Most MSN applications rely on accurate location information, which makes location an important asset for nodes throughout the network. One of the examples of location confidentiality approaches was proposed by Gongjun *et al.* [54]. The aim of this work is to provide a set of location information security mechanisms to meet the requirements of the confidentiality, integrity, and availability in vehicular ad hoc networks. To do that, a set of location security mechanisms such as onboard radar devices and GPS has been integrated. The proposed method is capable of enhancing the availability of location information by selecting and maintaining stable routing paths.

An interesting example of location attacks called *location cheating attack* [55] was proposed. It has the ability to pass the current location verification mechanism in MSN services. Fig. 8 portrays an example of location cheating which targets a foursquare server in a mobile environment. This attack has a simple but efficient mechanism. Basically, the location cheater creates a fake location and forces the server to reveal that location. As soon as the server connects to the falsified location, the connection between the true location and the server is blocked. The authors categorized possible solutions against it and provided insights into the defending mechanisms. In this paper, location verification techniques, namely, distance bounding, address mapping, and venue side location verification, were employed to excess resistance. To mitigate the threats, however, access control for crawling and hiding information from profiles was recognized.

*C. Intrusion Detection*

Network intrusion is an important attack in which a malicious user gains unauthorized access to a protected network. To tackle such attacks, detection and prevention methods have been used as proactive security measures, rather than waiting for an actual intrusion to occur. No matter how many intrusion prevention measures are inserted in a network, there are always some weak links that one could exploit to break in. IDSs monitor a network and produce reports about any malicious activities or policy violations throughout the network. The reports can be kept in a data set to be utilized for submitting brand new security policies. Some IDSs are equipped with alerting mechanisms to produce real-time reports about any intrusions. Some others try to stop an intrusion attempt by dropping the malicious packets, resetting the connection, and blocking the traffic from the offenders. Fig. 9 illustrates an example of IDS that monitors the mobile network and catches an intruder who gained an unauthorized access to the mobile network, bypassing security authorization and access policies.

Unlike wired networks, where extra protection can be placed on routers and gateways, an adversary node could paralyze the entire wireless network by disseminating false routing information. Such false routing information could result in messages from all nodes being fed to the compromised node. From a system security perspective, there have been efforts on designing IDSs [56] for pervasive computing. IDSs can be broadly divided into three classes: misuse detection (often also called signature-based detection), anomaly detection, and specification-based detection. Anomaly detection tries to look for behaviors that deviate from normal and expected behaviors. Statistical techniques are used to infer an anomaly; however, this can lead to false positives.

The internal of an intrusion detection agent for mobile networks [57] can be structured into six pieces. The data collection module gathers local audit traces and activity logs, which are utilized by the local detection engine to detect local anomaly. The cooperative detection engine fulfills the necessity of broader data sets or intrusion detection agents' collaborations. Intrusion response actions are provided by both the local response and global response modules. The local response module triggers actions local to this mobile node while the global one coordinates actions among neighboring nodes. Finally, there exists a secure communication module, which provides a secure communication channel among intrusion detection agents. Fig. 10 demonstrates IDS architecture for mobile networks along with its interior design.

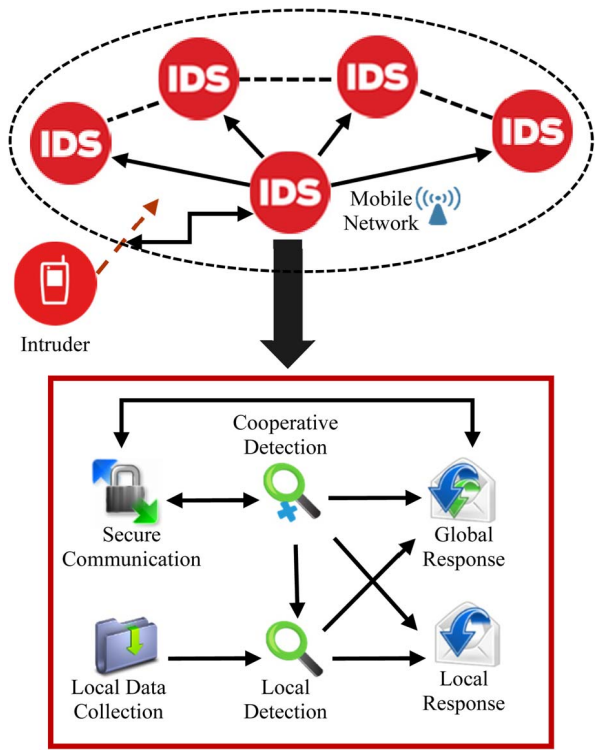

Fig. 10. Exterior and the interior design of intrusion detection agents and their cooperation in mobile network environment.

An interesting project in this area which fits in well with MSNs was proposed by Wang *et al.* [58]. It is a cooperative intrusion detection architecture, which relies on a detection engine to utilize social network analysis methods. In this architecture, each node deploys an intrusion detection engine that performs detections using audit data received from its "*ego*" network. The deployed engines operate similarly to anomaly detection, but they utilize social relations as metrics of interest. They require less computational overhead compared to standard anomaly detection engines. This architecture is composed of three modules: data preprocessing module to collect and preprocess the audit data, social analysis module to perform intrusion detection, and response module to integrate local and global intrusion alerts. Table II portrays our classification of security and discusses related approaches briefly.

## V. Privacy

Network privacy is the ability of an individual or group to reveal its information selectively and remain unnoticed or unidentified over the network. It encompasses anonymity, information blurring, and furtiveness of exchanged messages between intermediate nodes. The level of privacy in MSNs depends on the application, the point of view (sender, receiver, intermediate node, and outside observer), and the level of the trust between entities. In the case of context-based forwarding, the context of the message is directly linked to the profile of the destination and is considered as private. The context should therefore be protected (encrypted) while the information about the shared context should be revealed. This raises the problem of computation on encrypted data. In the case of content-based forwarding, preserving the privacy of users mainly consists of protecting their interests. In this case, users want to receive content corresponding to their interests without revealing them. However, user privacy and forwarding present conflicting requirements. The first requires the encryption of the interests, while the second requires access to the filters. This raises another problem of computation on encrypted data. To clarify privacy challenges in MSNs, we have partitioned them into the following different classes according to their behavior: 1) *obfuscation*; 2) *fairness encouragement*; 3) *private matching*; 4) *location privacy*; and 5) *communication privacy*.

### A. Obfuscation

Privacy preferences are generally specified to govern context exchange among nodes in ubiquitous environments. Aside from who has rights to see what information, a user's privacy preference could also designate who has rights to have what obfuscated information. By obfuscation, people are able to present their private information in a coarser granularity, or simply in a falsified manner, depending on the specific situations. In other words, obfuscation is a form of data masking where data are purposely scrambled to prevent unauthorized access to sensitive materials. This form of encryption results in unintelligible or confusing data.

A popular and traditional mechanism for privacy enhancement and anonymous communication over a network is onion routing [59]. Using this mechanism, messages are repeatedly encrypted and routed through a group of collaborating nodes to prevent the intermediary nodes from knowing the origin, destination, and content of the message. Like someone peeling an onion, each onion router removes a layer of encryption to uncover routing instructions and sends the message to the next router where this is repeated. To combine onion routing and multicast routing in mobile networks, Aad *et al.* [60] introduced methods to improve anonymity by using bloom filters to compress and obscure a packet's routing list.

Aside from onion-based routing schemes, one of the important works around privacy is a method called none of your business (NOYB) [61] which uses obfuscation to enhance privacy in social network sites. It combines users' information from different sites to prevent an attacker from profiling an individual user. In order to achieve privacy, this method is equipped with several mechanisms such as the marginal distribution of the ciphertext, atom compartmentalization, steganography, public dictionary, random nonce, standard ciphers, and communication across different channels. Later, NOYB was extended by "√" – if the scheme satisfies the property, "×" if not a defense strategy called *hide-and-lie* [62] which obfuscates users' interests in an opportunistic publish–subscribe application. It simply prevents attackers from identifying a user's specific interests. Using this strategy, the success probability of an attacker can be equalized to the success probability of the simple guessing. Furthermore, in some scenarios, the hide-and-lie strategy increases the message delivery ratio.

TABLE II
DEMONSTRATION OF SOME RECENT SAFETY SCHEMES IN SECURITY CATEGORY IN MSNS

| Scheme | Functionality | access control | | | confidentiality | | intrusion detection |
|---|---|---|---|---|---|---|---|
| | | group based | attribute based | policy based | identity | location | |
| Con. Ctxt-bsd & Epdc [42] | Prevents data from being accessed by different groups, using multiple levels of cryptography. | √ | × | × | × | × | × |
| Practical Security [43] | Provides authenticated and encrypted communications that can only be decrypted by privileged users with valid node ID. | √ | × | × | × | × | × |
| SSH [44] | Grants each patient with a pseudo-ID and its private key to make mutual authentication. | × | √ | × | × | × | × |
| Revocation [45] | Enables a trusted authority to monitor the data decryption using an attribute-based encryption. | × | √ | × | × | × | × |
| Dyn. P-Based for P2P [46] | Allows a group to switch between policies over time so peers can collaboratively modify policies. | × | × | √ | × | × | × |
| P-B AccCon MbScEc [47] | Allows the representation of expressive policies based on users' social interactions while keeping policy model compatibility. | × | × | √ | × | × | × |
| ID Cryp 4 end2end [50] | Allows a source to derive the destination public key from some associated identity strings, using identity-based cryptography. | × | × | × | √ | × | × |
| Dynamic virtual digraph [51] | Distributes public key based on two-channel cryptography by extending graph theory and distinguishing between owners and carriers. | × | × | × | √ | × | × |
| Cloud [52] | Forces ambiguity and intelligence resistant search functionality by hiding individuality of hosts within groups of semantically close network. | × | × | × | √ | × | × |
| SSA [53] | Exchanges an encrypted nonce and combines it to a verified user location without any need for adjusting user privacy and security. | × | × | × | √ | × | × |
| LocSec for VNETs [54] | Provides a set of location information security mechanisms to meet the requirements of the confidentiality, integrity, and availability. | × | × | × | × | √ | × |
| Location Cheating [55] | Provides location verification techniques as defense, and uses access control and information hiding for threats mitigation. | × | √ | × | × | √ | × |
| IDS, SocNet AnalMd [58] | Places intrusion detection engine in every node to audit data received from its "ego" network using social network analysis methods. | × | × | × | × | × | √ |

Another work around obfuscation-based privacy in social networks was presented in [63]. This work studies the feasibility of perceivable social networks through the comparison of an anonymous data set to another available social network data set. One problem with this method is that, to randomize data, it is necessary to keep the statistics close to the origin, which will reveal the hidden data themselves. The second problem is that this method focuses on one-time releases. In other words, the republication of dynamic social network data has not been considered in this method.

In order to resolve the first problem and keep the data out of attackers' access, Wondracek *et al.* [64] suggested another solution. They proved that it is also possible to reveal social network data if group membership information is public. Later, an approach consisting of two complementary methods on this subject was presented by Parris *et al.* [65]. These methods enhance privacy in social network routing by obfuscating the friends' lists in order to inform routing decisions. To do this, one-way hashing technique is used which is independent from any kind of key management schemes. Utilizing three real-world data sets, this work evaluates the proposed methods and shows that it is feasible to use such methods without any reduction in routing performance.

In order to resolve the second problem, the authors in [66] show that, by utilizing correlations between sequential releases, the adversary can achieve high precision in the deanonymization of the released data. It lets enemies suppress the uncertainty of reidentifying each release separately and synthesize the results afterward. In addition, this work suggests a combination of structural knowledge with node attributes to compromise graph modification-based defenses.

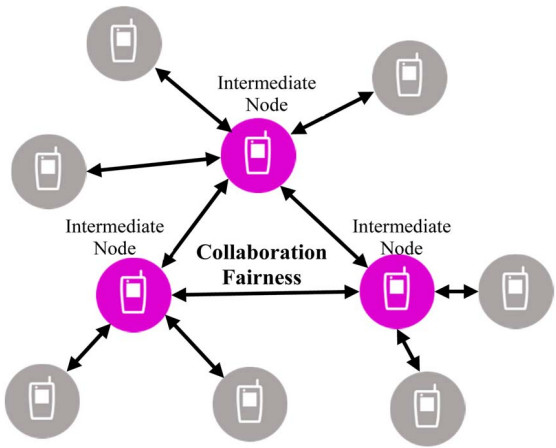

Fig. 11. Example of collaboration fairness in a mobile network where intermediate nodes try to take equal responsibility in message delivery.

## B. Fairness Encouragement

The concept of fairness is another fundamental issue in network privacy which generally covers methods of encouraging users to collaborate in a network. In order to clarify, we have categorized fairness into two major classes: *collaboration fairness* and *social fairness*. The next two sections discuss these classes in detail.

*1) Collaboration Fairness:* The first class of fairness, which we call collaboration fairness, encompasses fair contribution in message forwarding or longer network settlement. Providing fairness in this case is crucial since the unfair treatment of users is considered as a disadvantage to the participation in the communication process. However, an unfair communication between nodes causes a heavy congestion in a particular collaborative node. This node can be a dangerous high potential point for a malicious node to intrude into the network and gain an unauthorized access to valuable resources. Fig. 11 displays a state of collaboration fairness in which all intermediate nodes are coerced to cooperate and share their communication path in a relatively equal manner.

Regarding collaboration fairness in OppNets, higher ranked nodes typically take the responsibility of carrying the largest burden in message delivery actions, which creates a high potential of dissatisfaction among them. An absolute fair treatment of users, however, causes significant end-to-end delay and message delivery performance degradation. In other words, there is a tradeoff between fairness and efficiency since fairness schemes typically reduce the channel utilization.

It is questionable whether certain fairness schemes have a positive or negative impact on the quality of service (QoS). Dramitinos *et al.* [67] introduced a three-tier utility-base framework, based on history-dependent utility functions and QoS parameters. The aim of their approach is to quantify the satisfaction of the users from the way that their services are allocated. Bandwidth can be used as a holistic performance evaluation tool of these fairness schemes. Another project of this type of fairness is a real-trace-driven approach called FOG [68]. This approach basically studies and analyzes the tradeoff between the efficiency and fairness of rank-based forwarding techniques in OppNets. Using local information, it ensures efficiency–fairness tradeoff and guarantees relative equality in the distribution of resource usage among neighboring nodes. In addition, it keeps the success rate and cost performance near optimal.

*2) Social Fairness:* As users' interests and identification information are considered valuable resources in MSNs, the direct revelation of identity information to unknown users may result in an unfair situation when other sides of the identification process misbehave. This is the second class of fairness, which we call social fairness. This type of fairness focuses on the fair allocation of resources, specifically identity information revelation. As an essential factor of the wide acceptance of applications in distributed systems, fairness has thoroughly been researched. However, in emerging MSNs, establishing this type of fairness brings about a new set of challenges [69], including the following: distinguishing between an unfair cheating behavior and an unpredictable disconnection, information traceability and identifiability over the untrustworthy and short-term neighbors, and time sensitivity of location-based services.

In order to ensure perfect fairness, a trusted third party (TTP) had to be involved. Hence, an approach called gradual exchange protocol (GEP) [70] was presented to achieve fairness without TTP involvement by discouraging cheating behaviors. This protocol allows two users to reveal part of a secret to each other in iterations until they both have access to each other's entire secrets. Using the basics of GEP, years later, a non-TTP protocol for MSNs called fine-grained identification [71] was designed. This protocol provides confidentiality, weak linkage, and real-time fairness. For this protocol, identification is carried out by an iterative identification information exchange process where participants must disclose part of their identification information to each other. The process terminates whenever one of them fails to do so. In this way, if a user loses part of his/her identification information to another user, they must have obtained an approximately equal amount of identification information.

## C. Private Matching

Matchmaking is the key component of MSNs. It notifies users of possible candidates for partnership, based on some criteria such as shared interests. Personal profile disclosure, which is shown in Fig. 12, is an example of matchmaking that flusters users in proximity. The matchmaking process must happen before making decision for any interaction because attackers are in direct correlation with the personal profiles. This situation highlights the necessity for private matching to let two users conceal their personal profiles while in connection. Based on their mechanisms, private matching schemes can be identified in three subgroups, including *secret sharing*, *coarse-grained schemes*, and *fine-grained schemes*. These subgroups are described in the next three sections.

*1) Secret Sharing:* The first set of secret sharing approaches adopted homomorphic encryption. Apparently, the first attempt to implement such encryption dates back to FNP [72]. Using this scheme, a client and a server compute the intersection of their sets while the client gets the result and the server learns nothing. In other words, FNP takes advantage of homomorphic encryption to represent the client's input obliviously. Later, in [73], a complimentary scheme was proposed. This scheme not only enables set intersection, union cardinality, and overthreshold operations but also extends FNP to multiple players. This approach was later improved by Sang *et al.* [74] in which they considerably reduced computation and communication costs.

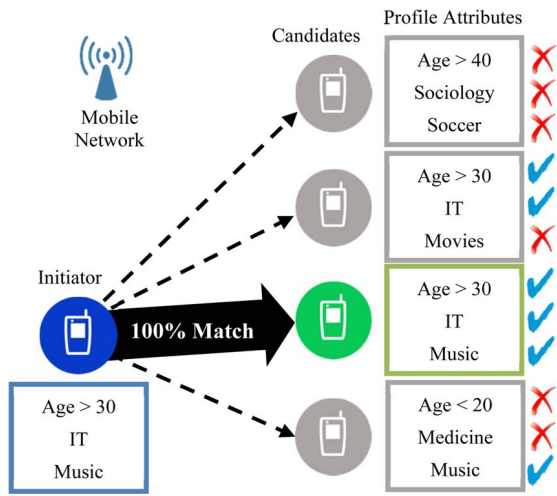

Fig. 12. Example of profile matchmaking where factors like age, university major, and interests have been considered as profile information.

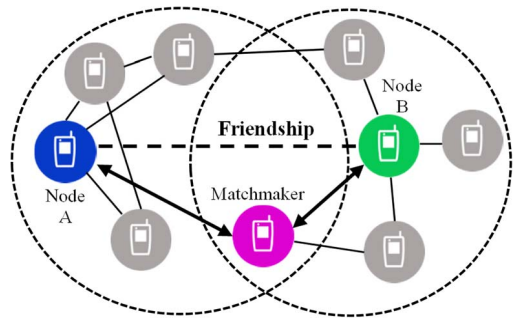

Fig. 13. Example of a friend-of-friend detection scheme in MSNs showing a matchmaking scenario where an intermediate node, which is in interconnection with different nodes and shares their profile information, tries to maintain friendship between them.

The problem with all these works is that, due to the strong dependence on homomorphic encryption and disability to implement linear computational complexity, almost none of them are practical enough to be applied to MSNs. To make such schemes compatible, some other lines of work pursue information theoretical security and try to follow secret sharing techniques. An important work in this area was proposed in [75]. This work uses a secret sharing scheme to provide a distributed solution of the FNP. In this scheme, the authors use a polynomial to represent one party's data set as in FNP and then distribute the polynomial to multiple servers. They extend their solution to the distributed set intersection and the cardinality of the intersection. Another example of secret sharing-based private matching was presented in [76]. This work suggests an unconditional secure multiparty set intersection scheme in which the inputs are shared among all parties using threshold secret sharing. Computations are done on those shares to obtain the shares of the outputs. Later in [77], the authors utilize secret sharing to compute the dot product securely through trusted third parties.

*2) Coarse-Grained Schemes:* The basic aim of coarse-grained private matching schemes is to match two users according to the privacy-preserving computation of the intersection (cardinality) of their attribute sets. Almost all these schemes implicitly assume that each user's personal profile consists of multiple public sets of characteristics derived from a public set of attributes. FindU [78] was the first coarse-grained privacy-preserving personal profile matching scheme for MSNs. It fulfills the primary privacy demand for a personal profile. An initiating user can find best matches according to their desired attributes while the actual set of matching attributes between the initiating user and any other user is hidden from all participants. FindU also contains different levels of user privacy. While leveraging secure multiparty computation techniques, it defines protocols to realize increasing levels of user privacy protection.

Some coarse-grained approaches focus on friendship discovery, using friend-of-friend detection algorithms. These algorithms mainly define an intermediate node called *matchmaker* that is responsible to establish interconnections between nodes, which have similar interests and with which it is in contact (see Fig. 13).

An example of mobile social networking platforms that implement the friend-of-friend detection algorithm is a privacy-preserving personal profile matching scheme called VENETA [79]. Rather than only exploiting information about the users of the system, the method relies on real friends and adequately addresses the arising privacy issues. It makes use of some notations, including commutative and homomorphic encryption, and passive and active adversaries to exhibit features like contact matching, decentralized messaging, server bound messaging, and user location tracking.

Later, Wei *et al.* [80] presented a piece of work, which focuses on developing some techniques and protocols to compute social proximity between two users in order to discover potential friends. It, however, identifies potential attacks against friend discovery to develop a solution for secure proximity estimation with privacy and variability considerations. In this way, their proposed approach does not match two users using the cardinality of their attribute sets. Instead, they proposed using the social proximity between two users as the matching metric, which measures the distance between their social coordinates with each being a vector precomputed by a trusted central server to represent the location of a user in an MSN.

To expand matchmaking protocols for MSNs, a privacy-preserving matchmaking protocol [81] has been proposed. It lets a potentially malicious user learn only the shared common interests with a nearby user. This protocol is distributed and does not require a trusted server to track users or be involved in any matchmaking operation. The mentioned mechanism offers the capability to defy against passive and active attacks like the user's interest exploration, impersonation, and eavesdropping.

*3) Fine-Grained Schemes:* Generally, the schemes based on coarse-grained private matching are unable to further differentiate users with the same attributes. To deal with this problem, fine-grained private matching protocols were proposed [82], which enable two users to perform profile matching without any need for information disclosures. In contrast to coarse-grained private matching schemes, these protocols allow finer differentiation between users and support a wide range of matching metrics at different privacy levels. The first protocol is for the first level distance, which is the sum of the absolute difference in each attribute. The second protocol is a threshold-based protocol. It is based on this distance too, in which two users can determine whether the distance between their profiles is smaller than some personally chosen threshold. The third protocol is based on the maximum distance, which is the maximum difference among attributes.

### D. Location Privacy

Social networks offer different types of location-based services such as photo sharing, friend tracking, and check-ins. However, to be able to deliver such services, users' locations have to be revealed, which causes privacy threats. Location privacy gives individuals or group users the ability to seclude or reveal their location selectively. The following four sections discuss different types of schemes designed to maintain this kind of privacy, including *obfuscation-based schemes*, *social-based schemes*, *dynamic pseudonymity*, and *key anonymity*.

*1) Obfuscation-Based Schemes:* Obfuscation was probably the first scheme to be employed to achieve location privacy. One of the first attempts in this area was proposed by Duckham and Kulik [83]. They use obfuscation in a formal framework to protect location privacy within a pervasive computing environment. The proposed framework provides a computationally efficient mechanism for balancing a user's need for high-quality information services against the user's need for location privacy. Negotiation is used to ensure that a location-based service provider receives only the information that it needs to know in order to provide a service of satisfactory quality.

The authors in [84] propose a different obfuscation technique to protect the location privacy of users. They present various obfuscation operators by changing their location information. In addition, they introduce an adversary model and provide an analysis of the proposed obfuscation operators to evaluate their robustness against adversaries.

Another example of obfuscation-based location privacy approaches is social-based location privacy protocol [85]. It offers location privacy through a request/reply location obfuscation technique. This protocol uses the nodes' own social network to drive the forwarding heuristic and utilizes social ties between nodes to ensure k-anonymity. This can result in a noticeable improvement in location privacy for users accessing location-based services.

*2) Social-Based Schemes:* Location privacy is considered a big concern in social networks. This is laid on some of the key features of these types of networks such as information sharing and content distributions. A fundamental research in this area has been done in [86] where privacy aspects in geosocial networks (GeoSNs) are studied. These types of social networks provide context-aware services that help associate location with users and content. In this paper, privacy aspects were classified in four categories (including location, absence, colocation, and identity privacy), and possible means of protecting privacy in these circumstances were described.

Another work was proposed by Li *et al.* [87], which addresses the location privacy issue for the nearby friend alert service, a common and fundamental service in mobile GeoSNs. In this paper, the grid-and-hashing paradigm was adopted, and an optimal grid overlay and multilevel grids were developed to increase the detection accuracy while saving the wireless bandwidth. Based on these techniques, the client-side location update scheme and the server-side update handling procedure for continuous proximity detection were devised.

In [88], the authors demonstrate a new type of user privacy attack and propose a solution for it. They used a fake location reporting technique to prevent an enemy from combining the location and friendship information found in MSNs. This solution does not require any additional trusted third party deployment and can enhance location privacy.

A common problem with most of location privacy-preserving approaches in MSNs is that the data-forwarding process can be interrupted or even disabled when the privacy preservations of users are applied. This is because users become unrecognizable to each other and the social ties and interactions are no longer traceable to facilitate cooperative data forwarding. Another problem with these approaches is that, to apply user cooperation, an intrusion must be exerted on user privacy. In order to solve such problems, social morality [89] was proposed. It is a protocol suite which achieves both privacy preservation and cooperative data forwarding in three steps. The first step is to notify a user's anonymized mobility information to the public using a privacy-preserving route-based authentication scheme. The second step measures the proximity of the user's mobility information to a specific packet's destination and evaluates the user's forwarding capacity for the packet. The third step determines the optimal data-forwarding strategy according to morality level and payoff, using a game-theoretical approach.

*3) Dynamic Pseudonymity:* Anonymity can be provided via the frequent changing of pseudonyms, in order to make it difficult for adversaries to detect a user's movement, information, location, etc. This technique has been widely adopted to provide location privacy. One of the finest approaches in this area was proposed by Magkos *et al.* [90]. In this paper, a distributed scheme which deals with both security and historical privacy in MSNs is suggested. In other words, it establishes access control while protecting the privacy of a user in both sporadic and continuous queries. To maintain security, this scheme employs a hybrid network architecture which gives users an ability to communicate with a location-based service provider through a network operator. Using this architecture, users are enabled to create wireless ad hoc networks with other users in order to obtain privacy against an adversary that performs traffic analysis. To maintain historical privacy, this scheme adopts the generic approach of using multiple pseudonyms that are changed frequently. Messages are not sent directly to the cellular operator but are distributed among mobile neighbors, so they can re-encrypt the messages before sending them to the location-based service provider via the cellular operator. This makes messages untraceable against traffic analysis attacks.

Another important work regarding dynamic pseudonymity is a privacy-preserving location proof updating system for location-based services called APPLAUS [91]. It utilizes colocated Bluetooth which enables mobile devices to mutually generate location proofs and update to a location proof server. This work contains a user-centric location privacy model to evaluate user location privacy levels in real time and gives users the ability to accept a location proof exchange request based on their location privacy levels. In this method, mobile devices use periodically changed pseudonyms to protect source location privacy.

However, APPLAUS could not resist collusion attacks and could only detect them. Regarding the fact that this method uses pseudonyms to keep contact patterns of nodes, if the pseudonyms are changed in an improper time or location, the solution offered in APPLAUS may become invalid. This can simply be the source of collusion attacks. To cope with this issue, Rongxing *et al.* presented an effective pseudonym changing at social spots (PCS) [92] for vehicular ad hoc networks. They introduce social spots as the place where several vehicles may gather and try to make vehicles change their pseudonyms at some highly social spots. Then, using PCS strategy, vehicles are able to intelligently change their pseudonyms at the right moment and place. Later, the authors of APPLAUS expanded their method [93] by incorporating two new approaches for outlier detection, including betweenness ranking-based approach and correlation clustering-based approach. These new features make APPLAUS extremely resilient against collusion attacks.

*4) Key Anonymity:* An important challenge in the wide deployment of location-based services is to provide safeguards for the location privacy of mobile clients against vulnerabilities for abuse. In order to achieve such a goal, key anonymity schemes have been widely deployed. For example, the authors in [94] develop a personalized location anonymization model and a suite of location perturbation algorithms to protect location privacy in the deployment of location-based services. This architecture takes advantage of a flexible privacy personalization framework to support location k-anonymity for a wide range of users with context-sensitive privacy requirements. The proposed framework enables each node to specify the minimum level of anonymity that it desires and the maximum temporal and spatial tolerances that it is willing to accept when requesting k-anonymity-preserving location-based services. The model is able to be run by the anonymity server on a trusted platform and can perform location anonymization on identity removal and spatiotemporal cloaking of the location information.

Traditional approaches to K-anonymity guarantee privacy over publicly released data sets with specified quasi-identifiers. However, due to the fact that common public releases of personal data are done through social networks and their application program interface (API), these approaches are barely responsible for today's needs. In other words, k-anonymity in social networks does not allow clear assumptions about quasi-identifiers, which makes traditional approaches impossible to be responsible for their privacy needs. In order to address this problem, Social-K [95] suggests a new definition of k-anonymity. In this definition, social networks guarantee privacy in real time to users of their API. In order to achieve privacy while improving the key update efficiency of location-based services in vehicular ad hoc networks, a dynamic privacy-preserving key management scheme called DIKE [96] was proposed. It uses a particular type of privacy-preserving authentication technique that not only provides the vehicle user's anonymous authentication but also enables double-registration detection as well. DIKE updates keys by dividing the session of a location-based service into several time slots so that each time slot holds a different session key. When no vehicle user departs from the service session, each joined user can use a one-way hash function to autonomously update the new session key for achieving forward secrecy. In addition, this scheme integrates a dynamic threshold technique to achieve the session key's backward secrecy.

*E. Communication Privacy*

Preserving an individual's privacy when communicating is another issue in every type of network. In order to give proper solutions, different privacy-enhancing technologies have been developed. That includes technologies like cryptography, authentication, and digital signatures. These technologies have various algorithms and protocols, which are used to a large extent in computer networks.

One example of the approaches in this area is PEACE [97]. It is a privacy-enhanced security framework consisting of authentication and key agreement protocols for wireless mesh networks (WMNs). These types of networks contain nodes with the ability to both disseminate their own data and propagate the data in the network. PEACE enforces strict user access control to cope with both free riders and malicious users. In addition, it offers user privacy protection against both adversaries and various other network entities.

Another example is the work proposed in [98]. This work presents a hybrid communication protocol to ensure mobile users' anonymity against various adversaries. It exploits the capability of handheld devices to connect to both Wi-Fi and cellular networks. The authors consider all parties that can intercept communications between a mobile user and a server as potential privacy threats. In addition, they describe how a micropayment scheme that suits their mobile scenario can provide incentives for peers to collaborate in the protocol.

To conclude, some of recent schemes in MSN privacy are discussed and highlighted in Table III.

## VI. DISCUSSION AND FUTURE DIRECTIONS

MSNs face various safety issues and challenges from different disciplines such as trust, security, and privacy. Despite extensive research on MSNs, focusing on the aforementioned aspects, there are still some questions left unanswered. In the remainder of this section, we go one step further by presenting some future research directions of safety challenges, which brings new vision into the horizon of safety in MSN research.

*A. Trust*

One of the capital assets in MSN safety is trust establishment and maintenance which has gained immense considerations, particularly in recent years. Widespread approaches such as that in [13] exist which aim to present ways to have smooth levels of trust between relations as in real-world social networks.

TABLE III
DEMONSTRATION OF SOME RECENT SAFETY SCHEMES IN PRIVACY CATEGORY IN MSNs

| Schemes | Functionality | obfuscation | collaboration fairness | social fairness | secret sharing | coarse grained | fine grained | obfuscation based | Social based | dynamic pseudonym | key anonymity | communication privacy |
|---|---|---|---|---|---|---|---|---|---|---|---|---|
| | | | fairness encourage | | Private matching | | | location privacy | | | | |
| PacketCoding[60] | Utilizes bloom filters to compress and obscure a packet's routing list. | √ | × | × | × | × | × | × | × | × | × | × |
| NOYB [61] | Prevents attackers from profiling users by combining their information from different social sites. | √ | × | × | × | × | × | × | × | × | × | × |
| hide-and-lie [62] | Prevents attackers from identifying user interests by obfuscating them in opportunistic publish-subscribe application | √ | × | × | × | × | × | × | × | × | × | × |
| Privacy-enhcd Routing [65] | Obfuscates the friends' lists used to inform routing decisions, by using one-way hashing. | √ | × | × | × | × | × | × | × | × | × | × |
| D-anon. Dyn. SocNet [66] | Shows utilizing correlations between sequential releases, an enemy can achieve high precision to in de-anonymize data. | √ | × | × | × | × | × | × | × | × | × | × |
| Prf eval for Fair Slt. [67] | Measure the service bandwidth satisfaction among users using history-dependent utility functions and QoS parameters. | × | √ | × | × | × | × | × | × | × | × | × |
| FOG [72] | Ensures efficiency-fairness tradeoff and guarantees relative equality in the distribution of resource usage among neighbors. | × | √ | × | × | × | × | × | × | × | × | × |
| F.Gd.ID with R.T. Fair [71] | Makes nodes disclose part of their ID information and exchanges the rest in an iterative process. | × | × | √ | × | × | × | × | × | × | × | × |
| FNP [68] | Makes a client and a server compute the intersection of their sets while the client gets the result and server learns nothing. | × | × | × | √ | × | × | × | × | × | × | × |
| Mlt-party set intersec. [76] | Forces nodes to share inputs among all parties and computes only those shares that obtain the shares of the outputs. | × | × | × | √ | × | × | × | × | × | × | × |
| Cryp Prv data mining [77] | Computes dot product securely, though trusted third parties. | × | × | × | √ | × | × | × | × | × | × | × |
| FindU [78] | Hides matching attributes by leveraging secure computation techniques to realize levels of user privacy protection. | × | × | × | × | √ | × | × | × | × | × | × |
| VENETA [79] | Exhibits contact matching, decentralized/ server bound messaging and user location tracking. | × | × | × | × | √ | × | × | × | × | × | × |
| Secure Friend Discov. [80] | Discover the location of potential friends and measure the distance between their social coordinates. | × | × | × | × | √ | × | × | × | × | × | × |
| PP-Match VS Mal [81] | Prevents malignity in matchmaking by sharing common interests in vicinity and omitting track-taking of trusted servers. | × | × | × | × | √ | × | × | × | × | × | × |
| Fine-GR matching [82] | Enables users to match without any need for information disclosure using differentiation metrics at privacy levels. | × | × | × | × | × | √ | × | × | × | × | × |
| Formal Obf [83] | A framework in which obfuscated location-based services are defined. | × | × | × | × | × | × | √ | × | × | × | × |
| SLPD [85] | Offers location privacy through a request/reply location obfuscation. | × | × | × | × | × | × | √ | × | × | √ | × |
| Enhancing MSN Privacy [88] | Reports a fake location to prevent an enemy from combining the location and friendship information. | × | × | × | × | × | × | × | √ | × | × | × |
| Social morality [89] | A three-step protocol suite to achieve both privacy preservation and cooperative data forwarding. | × | × | × | × | × | × | × | √ | × | × | × |
| Distributed PrivPreserv [90] | An access control system to protect the privacy of a user in both sporadic and continuous queries. | × | × | × | × | × | × | × | × | √ | × | × |
| APPLAUS [91],[93] | Makes Bluetooth enabled mobile devices mutually generate location proofs, and uploads to the location proof server. | × | × | × | × | × | × | √ | × | √ | × | × |
| Social-k [95] | A new definition of K-anonymity in which social networks could provide privacy guarantees to users of their API. | × | × | × | × | × | × | × | × | × | √ | × |
| DIKE [96] | A dynamic privacy-preserving key management and LBS session key update scheme. | × | × | × | × | × | × | × | × | × | √ | × |
| PEACE [97] | A framework for WMNs which enforces strict user access control to cope with both free riders and malicious users. | × | × | × | × | × | × | × | × | × | × | √ |
| User's Anonymity in MHNs [98] | Presents a hybrid communication protocol to ensure mobile users' anonymity against a various adversaries | × | × | × | × | × | × | × | × | × | × | √ |

"√" – if the scheme satisfies the property, "×" if not However, transitivity in certification chains and community detection define no barriers among social categories. This may not reflect reality accurately. Depending on the environment, the explicit and the implicit social trust should be weighed dynamically. In addition, an aging mechanism has to be applied to remove old or random encountered users and not end up with a clique resulting in a meaningless even structure. As an alternative, RSs can provide higher levels of trust such as taste similarity by comparing ratings, but they also have to be secured. Further research must investigate the interrelation between social trust and RSs.

Some other considerable future challenges regarding trust lie behind opportunistic sensing [25] which tends to integrate communication more closely with human behavior. This integration could lead to innovative applications that can sense the context of the user with a higher accuracy, thus providing a more personalized solution. For the next generation of sensing systems, an architecture based on the control and management of trustworthy social computing like socially aware network services is needed. Meanwhile, as trustworthiness for an application is enhanced, a simultaneous need for ensuring social informatics' consumption and processing will be required. To do this, the following questions must be answered: How do we determine the appropriate boundary for social informatics, and how do we quantitatively measure the value of social informatics? How do we balance the concern of privacy and anonymity by introducing a socially aware system? How can we make sure that we do not create any new or unexpected vulnerability? Which process must be taken into consideration in order to form a converging and collaborative decision about high-level security?

To design most of the protocols associated with safety, public-key management plays an important role. For a safe incentive scheme, any misbehaving or malicious nodes will pay the penalty of having their public-key certificates revoked. Even for those selfish nodes that run out of their credits, one possible punishment action is also revoking their certificates or reducing their class-of-service right by revising their certificates. However, a public-key revocation scheme [45] still represents a great challenge in MSNs because the nodes may suffer from delayed or frequent loss of connectivity to certificate revocation list servers and may lead to a lot of extra management costs. Schemes like SMART [33], which use traditional public-key-certificate-based cryptography as the basic cryptographic tool, are potential cases for further improvement. This can be done by adopting identity-based cryptography to recognize the efficient bundle authentication. This is a relatively new cryptographic method and is also a powerful alternative to traditional certificate-based cryptography.

Meanwhile, opportunistic communication and information dissemination in social networks are surprisingly robust toward the form of altruism distribution [35], largely due to their multiple forwarding paths. So far, altruism distribution has been considered static and steady. It would be interesting to study the altruism values resulting from gaming strategies to get feedback from prior delivery history. In many cases, users are not supposed to memorize their own previous rejection history for message delivery of other nodes while variations, like limiting the number of delivery requests for a single user to avoid excessive requests, could be studied. In addition, the power consumption of nodes and altruism effect on it would be another interesting issue to research. For example, one could measure how much power a node might save by choosing to be selfish compared with being only altruistic within its community.

The idea of leveraging social networking information enables node cooperation in mobile OppNets. In most papers on this topic like [38], data sets do not represent performance of the trust filters in large-scale networks. In other words, they do not indicate whether these trust filters can be efficiently estimated and implemented using a distributed algorithm running with local information at the nodes. However, a detailed mechanism for sharing and transmitting this social information upon which trust is based is crucial.

### B. Security

Security is a classic angle of network safety that has been massively discussed in the MSN literature. Albeit widespread research on this area, proposed schemes still need refinements or adjustment. Anonymous social-location-based architectures such as that in [53], which do not imply a specific implementation of any particular component, are examples of these schemes. They should be managed through a centralized infrastructure on the Internet or integrated into a P2P trust network. However, the way in which preferences are chosen and returned to the stationary component could be optimized for different metrics. This mechanism could be designed to provide k-anonymity for a set of users' encrypted identifier rather than just one at a time. This relates to a more general set theory problem. A set of social network information associated with a set of users should be chosen in a way that the set of preferences cannot map back to any one or any set of the users within some guarantee.

Another concept around security is cloud [52] which employs provision for vagueness and counterintelligence resistant investigation functionality. Although the influence on recovery potential has been ensured and a rich data replica and query language have been born, the secrecy and counterintelligence resistance procedures conceptualized are natural and can be used to novelize any unstructured overlay. Future research includes expanding the scrutiny to other modernized assault environments. This can be done for a Sybil attack where a spiteful host manages to fake multiple identities and assaults. It can also be applied for an Eclipse attack where a host is detached from other hosts and then assaulted. In addition, it is probable to employ certain characteristics of the prevalent network in order to safeguard itself against Sybil or similar assaults. To deal with the attacks like location cheating [55], the countermeasures against location cheating should be revised. Although several techniques for the security enhancement of location information have been suggested, finding better solutions to identify possible cheaters and maintain balance between the usability and the security still remain challenging.

The concept of IDSs has been vastly studied, and experimental results demonstrate a high rate of adaptation to MSN security. For instance, an anomaly detection approach like that in [57] can work well on different wireless ad hoc networks, including MSNs, but the question is how to detect an intrusion that relies on a single maneuver? For example, anomaly detection can be very effective against multiconnection-based port scan, DoS, and buffer-overflow attacks. This shows that there are some natural limits on detection capabilities, depending on which layer the data are collected. However, a few system parameters exist that may change the normal behavior tremendously. One of them is the mobility level; if the model is classified using values from another mobility level, the alarm rate can be much higher. Although this can be solved by randomizing the mobility level in the experiment, the current ns-2 code does not yet support this feature. To this end, schemes must be developed to cluster and classify the normal scenarios and build specific anomaly detection models for each type of normal scenarios.

### C. Privacy

One of the most controversial issues around privacy which has attracted attentions due to its novelty is fairness. Finding the most optimal tradeoff between fairness and efficiency is vital in MSNs, but unfortunately, not much has been done in this area. To quantify the satisfaction of the users from the way that their services are allocated, linear schemes like utility function scheme [67] exist. However, any linear combination of utilities is unable to ensure the ambiguity of social welfare values even if multiple types of service flow are used. Moreover, applying standard methods of experimental economics, such as the mean opinion score, can result in much more vigorous schemes to preserve fairness and maintain its balance with efficiency.

To preserve privacy in MSN applications, matchmaking protocols such as those in [82] have been broadly studied, but some questions, like how real people use matchmaking protocols to learn more about the usefulness of MSN application, still remain unanswered.

## VII. CONCLUSION

The concept of MSNs is a novel social communication paradigm that exploits opportunistic encounters between human-carried devices and social networks. Like any other emergent archetype of technology, MSNs demand time to be totally safe and immune. Having social aspects included, they encompass more complex and correlated challenging safety problems that make it difficult to suggest solutions and represent a clear classification on safety issues. This paper has aimed to provide an overall view of safety challenges in this young and exciting field, particularly from the perspective of trust, security, and privacy. To provide a comprehensive and precise investigation, each category was divided into different smaller subcategories. The trust-related issues were discussed in four categories, namely, malignity prevention, assorted trustworthiness, selfishness discouragement, and cooperation enforcement. The security-based challenges were deeply investigated in three groups, namely, access control, confidentiality, and intrusion detection. The privacy-engaged provinces were indicated and argued in three classes, namely, obfuscation, fairness encouragement, and private matching. Consequently, the major issues related to safety issues, lately discussed in the literature, were described. Finally, several major open research issues were discussed, and future research directions were outlined.

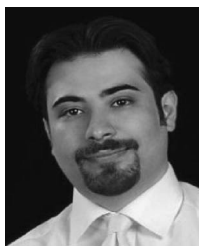

**Yashar Najaflou** received the B.Eng. degree in information technology from Payame Noor University, Tabriz, Iran, in 2012.

He is currently a Researcher with the School of Software, Dalian University of Technology, Dalian, China, working on mobile social networks (MSNs). He works mainly on artificial intelligence algorithms, game theory, and social network analysis to optimize security issues in MSNs.

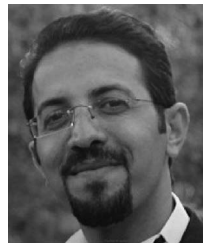

**Behrouz Jedari** received the B.Sc. and M.Sc. degrees from the Islamic Azad University, Qazvin, Iran, in 2006 and 2009, respectively. He is currently working toward the Ph.D. degree in the School of Software, Dalian University of Technology, Dalian, China.

His current research interests include delay tolerant networks and social network analysis with respect to data communication and protocol design for new and emerging areas such as mobile social networks.

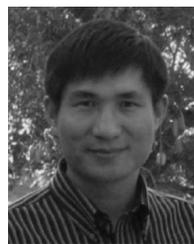

**Feng Xia** (M'07–SM'12) received the B.E. and Ph.D. degrees from Zhejiang University, Hangzhou, China, in 2001 and 2006, respectively.

He is an Associate Professor and the Ph.D. Supervisor in the School of Software, Dalian University of Technology, Dalian, China. He has authored/coauthored one book and over 140 scientific papers in international journals and conferences. His research interests include social computing, mobile computing, and cyber-physical systems.

Dr. Xia is a senior member of the IEEE Computer Society and the IEEE Systems, Man, and Cybernetics Society and a member of the Association for Computing Machinery (ACM) and ACM SIGMobile.

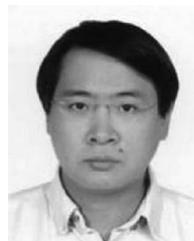

**Laurence T. Yang** received the B.E. degree in computer science and technology from Tsinghua University, Beijing, China, and the Ph.D. degree in computer science from the University of Victoria, Victoria, British Columbia, Canada.

He is currently a Professor with the School of Computer Science and Technology, Huazhong University of Science and Technology, Wuhan, China, and with the Department of Computer Science, St. Francis Xavier University, Antigonish, NS, Canada. His research interests include parallel and distributed computing and embedded and ubiquitous/pervasive computing. His research has been supported by the National Sciences and Engineering Research Council, and the Canada Foundation for Innovation.

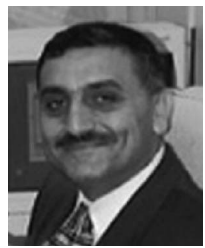

**Mohammad S. Obaidat** (S'85–M'86–SM'91–F'05) received the Ph.D. and M.S. degrees in computer engineering with a minor in computer science from Ohio State University, Columbus, Ohio, USA.

He is currently a Full Professor of computer science with Monmouth University, West Long Branch, New Jersey, USA. He has published over 15 books and over 560 refereed technical articles. He is the Editor-in-Chief of three scholarly journals and is also an Editor or Advisory Editor of numerous international journals.

Prof. Obaidat has chaired numerous international conferences and given numerous keynote speeches all over the world. He is a fellow of the Society for Modeling and Simulation International.